\documentclass[prl,twocolumn,showpacs,preprintnumbers,amsmath,amssymb,  superscriptaddress]{revtex4}

\usepackage{graphicx}
\usepackage{dcolumn}
\usepackage{bm}
\usepackage{epsfig}

\def\e{\begin{equation}}
\def\f{\end{equation}}
\def\%#1{\mbox{\boldmath $#1$}}
\def\=#1{\overline{\overline #1}}

\def\_#1{{\bf #1}}

\def\.{\cdot}

\def\##1{{\bf#1\mit}}
\def\Re{{\rm Re\mit}}

\def\lab#1{\label{eq:#1}}
\def\r#1{(\ref{eq:#1})}
\def\_#1{{\bf #1}}
\def\vec#1{{\bf #1}}

\begin{document}

\preprint{APS/123-QED}

\title{Strong spatial dispersion in wire media in the very large
wavelength limit}

\author{P.A. Belov} \affiliation{Radio Laboratory, Helsinki
University of Technology, P.O. Box 3000, FIN-02015 HUT, Finland}
\affiliation{Physics Department, St.\ Petersburg Institute of Fine
Mechanics and Optics, Sablinskaya 14, 197101, St. Petersburg,
Russia}
\author{R. Marqu\'es} \affiliation{Department of Electronics and
Electromagnetism, Facultad de F\'isica, University of Sevilla, Av.
Reina Mercedes s/n, 41012 Sevilla, Spain}
\author{S.I. Maslovski} \affiliation{Radio Laboratory, Helsinki
University of Technology, P.O. Box 3000, FIN-02015 HUT, Finland}
\author{I.S. Nefedov}
\affiliation{Radio Laboratory, Helsinki University of Technology,
P.O. Box 3000, FIN-02015 HUT, Finland} \affiliation{Institute of
Radio Engineering and Electronics, Russian Academy of Science,
Zelyonaya 38, 410019, Saratov, Russia}
\author{M. Silveirinha}
\affiliation{Instituto Superior T\'ecnico, Instituto de
Telecomunica\c{c}\~{o}es, Av. Rovisco Pais, 1049-001, Lisboa,
Portugal}
\author{C.R. Simovski}
\affiliation{Radio Laboratory, Helsinki University of Technology,
P.O. Box 3000, FIN-02015 HUT, Finland} \affiliation{Physics
Department, St.\ Petersburg Institute of Fine Mechanics and
Optics, Sablinskaya 14, 197101, St. Petersburg, Russia}
\author{S.A. Tretyakov}
\affiliation{Radio Laboratory, Helsinki University of Technology,
P.O. Box 3000, FIN-02015 HUT, Finland}

\date{\today}

\begin{abstract}
It is found that there exist composite media that exhibit strong spatial
dispersion even in the very large wavelength limit.
This follows from the study of lattices of ideally
conducting parallel thin wires (wire media).
In fact, our analysis reveals that the description of this medium by means of a
local dispersive uniaxial dielectric tensor is not complete,
leading to unphysical results for the propagation of
electromagnetic waves at any frequencies. Since non--local
constitutive relations have been usually considered in the past as
a second order approximation, meaningful in the short wavelength
limit, the aforementioned result presents a relevant theoretical
interest. In addition, since such wire media have been recently
used as a constituent of some discrete artificial media (or
metamaterials), the reported results open the question of the
relevance of the spatial dispersion in the characterization of
these artificial media.
\end{abstract}


\maketitle

\begin{figure}[h]
\centering \epsfig{file=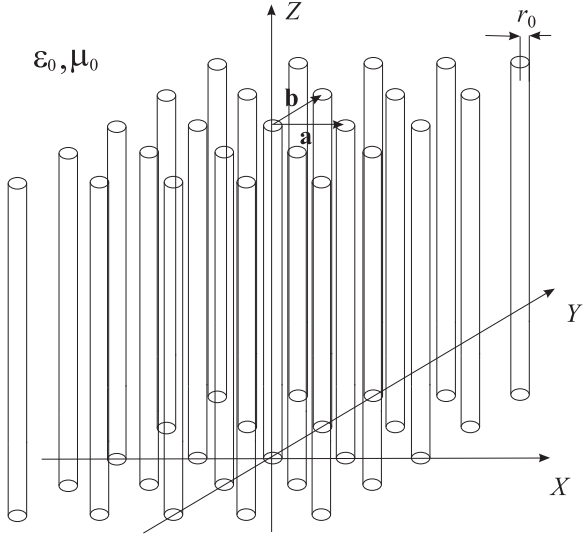, width=8cm} \caption{The
geometry of wire media: a lattice of parallel ideally conducting
thin wires.} \label{geom}
\end{figure}

Causality imposes that all material media must be dispersive. In
most cases this behavior results in local dispersive constitutive
relations -- i.e. in frequency dependent constitutive permittivity
and permeability tensors. Non--local dispersive behavior (i.e.
{\emph{spatial dispersion}), which results in constitutive
operators depending also on the spatial derivatives of the mean
fields (or, for plane electromagnetic waves, on the wavevector
components) is usually considered as a small effect,
meaningful in the short wavelength limit. Specifically, spatial
dispersion will always appear when the higher order terms in the
series expansion of the constitutive parameters in power series of
the dimensionless parameter $a/\lambda$ ($a$ is the lattice
constant of the crystal and $\lambda$ the wavelength inside the
medium) are not neglected \cite{Landau}. Thus, it is usually
assumed that non--local dispersive reations are only meaningful
when $\lambda$ approaches to $a$. The usually weak natural optical
activity of some materials is a well known example of the
application of this principle \cite{Landau}. When such principle
is translated to the analysis of discrete artificial media, also
called metamaterials, it would imply that non--local dispersive
constitutive relations are only expected to be a small
refinement of the local constitutive relations usually
considered. However, there is at least a counter example for this
assumption.

The parallel wire medium is a medium formed by a regular lattice
of ideally conducting wires with small radii compared to the
lattice periods and the wavelength, see Fig.~\ref{geom}. It has
been known in microwave applications for a long time
\cite{Brown,Rotmanps,King} as an artificial dielectric, also
called {\it rodded medium} and quasistatic models of the effective
permittivity are available \cite{StasMOTL}. Recently, some attention to wire media
has been paid also in optics (e.g. \cite{Pendryw,Pitarke}) and in
the realization of left--handed media) \cite{Veselago,Pendrylens}
as composite media made from lattices of long conducting wires and
split ring resonators \cite{Smithsr,Shelbysc,Smithnr}.} (see discussions corresponding to that 
in \cite{Utah}).
The electromagnetic reponse of the specific wire medium shown in
Fig.~\ref{geom} is analyzed in \cite{Rotmanps,Pitarke} following
different approaches. Both analysis are carried out for wave
propagation perpendicular to the wires and show that, for electric
field polarization along the wires, the medium is characterized
(if $a/\lambda\,,b/\lambda\ll 1$) by a frequency dependent
effective dielectric constant given by:
\e
\varepsilon=\varepsilon_0
\left(1-\frac{k_0^2}{k^2}\right)=\varepsilon_0
\left(1-\frac{\omega_0^2}{\omega^2}\right). \lab{epl} \f The constant
$\omega_0$ (the corresponding wavenumber
$k_0=\omega_0\sqrt{\varepsilon_0\mu_0}$) in \r{epl} plays the role
of an equivalent ``plasma frequency." Thus, this medium is often
called ``artificial plasma'' since the ideal (collisionless)
electron plasma is described by the same equivalent parameter. If
the wires are assumed to be very thin, so that their polarization
in the direction orthogonal to the wires can be neglected, the
effective permittivity for electric field polarization orthogonal
to the wires is $\varepsilon_0$.

The aforementioned analysis suggest that this wire medium could be
modeled as an uniaxial dielectric with the following \emph{local}
permittivity dyadic:
\e
\=\varepsilon=
\varepsilon\_z_0\_z_0+\varepsilon_0(\_x_0\_x_0+\_y_0\_y_0),\lab{uni}
\f whose  permittivity in the axial direction, $\varepsilon$,
would be given by \r{epl}.
However, it will be shown in the following that this \emph{naive}
hypothesis lead to unphysical results and must be substituted by a
non--local dispersive relation. In fact, assuming that the medium
can be described by the uniaxial dyadic \r{uni}, the dispersion
equation for extraordinary plane waves ($E_z\ne 0$) with the
wavevector $(q_x,q_y,q_z)^{\rm T}$ in this uniaxial dielectric
reads \cite{Ginzburg,Lindelldis}: \e \varepsilon_0
(q_x^2+q_y^2)=\varepsilon (k^2-q_z^2)\,, \lab{disuni} \f where
$k=\omega\sqrt{\varepsilon_0\mu_0}$ is the phase constant of the
host matrix. On the other hand, these extraordinary waves
correspond to the well known $TM$-(to $z$) set of modes, allowed
by the invariance of the boundary conditions along $z$. Thus, for
any extraordinary wave travelling with a phase constant $q_z$ along
the $z$-axis, the $E_z$ field must satisfy the Helmholtz equation
\e \left\{ \frac{\partial}{\partial x^2} +\frac{\partial}{\partial
y^2} + (k^2-q_z^2) \right\}
 E_z = 0\,, \lab{Helmholtz} \f with the boundary condition $E_z=0$ on
the wires. It is clear from this equation that any ``plane''
extraordinary wave must satisfy
\e
k(q_x,q_y,q_z)=\sqrt{k^2(q_x,q_y,0)+q_z^2}. \lab{total} \f This last
result is incompatible with \r{epl}--\r{disuni}, as can be easily
seen by substitution of \r{epl} into \r{disuni}. However, if we
choose
\e
\varepsilon(k,q_z)=\varepsilon_0
\left(1-\frac{k_0^2}{k^2-q_z^2}\right) \lab{eps} \f instead of
\r{epl}, then \r{uni} and \r{disuni} become compatible with
\r{total}, giving the following dispersion equation for the plane
wave: \e q^2\equiv q_x^2+q_y^2+q_z^2=k^2-k_0^2\,, \lab{qq} \f where
we have assumed that $q_z\ne k$ (the case with $q_z=k$ will be
analyzed at the end of this letter). The above rationale suggests
that the considered wire media still can be described by the
permittivity dyadic \r{uni}, but the axial permittivity
$\varepsilon$ must be a non--local parameter of the form \r{eps}.
The conventional expression \r{epl} would be only a particular
case of \r{eps}, valid for wave propagation in the $x-y$ plane.

The main difference between the local uniaxial model, \r{epl}, and
the proposed non--local model, \r{eps}, for the parallel wire
medium is that the non--local model predicts a stop band (at
frequencies below $\omega_0=k_0/\sqrt{\varepsilon_0\mu_0}$) for
extraordinary waves propagating along any direction in the media.
On the contrary, \r{epl}--\r{disuni} predict propagation of
extraordinary waves at any frequency provided
$q_z>k=\omega\sqrt{\mu_0\varepsilon_0}$. Thus, both models predict
qualitative very different behaviors, even near the cutoff
``plasma'' frequency, $\omega_0$, where $q^2\rightarrow 0$ (i.e.
$a/\lambda \rightarrow 0$). That is, the non--locality of the
proposed constitutive relations affects the electromagnetic
response of the medium even in the very large wavelength limit,
thus being important for any values of the $a/\lambda$ ratio
inside the medium. Other relevant differences between the
predictions of both models will be developed along this letter.

The rigorous proof of \r{eps} is based on the local field approach
which is described in detail in \cite{JEWAwm}. In \cite{JEWAwm}
the low frequency stop band of the wire medium has been analyzed,
as well as its high frequency band gap structure. This analysis
reveals that, in the thin wire medium (Fig.~\ref{geom}) and for
$q_z\ne k$, two sets of modes can propagate: ordinary (with
$E_z=0$) and extraordinary (with $E_z\ne 0$) waves. The ordinary
waves do not interact with the wires and propagate in the host
media. For extraordinary waves, an explicit dispersion equation
connecting the wave vector $\vec q=(q_x, q_y, q_z)^{\rm T}$ with
the wave number of the host isotropic matrix
$k=\omega\sqrt{\varepsilon_0\mu_0}$ has been derived in
\cite{JEWAwm}. It can be written as:
\e
\frac{1}{\pi}\log{\frac{b}{2\pi r_0}}+
\frac{1}{bk_x^{(0)}}
\frac{\sin k_x^{(0)}a}{\cos k_x^{(0)}a-\cos q_xa }+
\lab{disp}
\f
$$
\sum\limits_{n\ne 0}
\left( \frac{1}{bk_x^{(n)}}
\frac{\sin k_x^{(n)}a}{\cos k_x^{(n)}a-\cos q_xa }-\frac{1}{2\pi |n|}\right)=0.
$$
Here $k_x^{(n)}$ denotes the $x$-component of $n$-th Floquet mode wave vector:
\e
k_x^{(n)}=-j\sqrt{\left(q_y+\frac{2\pi n}{b}\right)^2+q_z^2-k^2},
\qquad \Re\{\sqrt{()}\}>0. \lab{kf} \f The other notations are clear
from Fig.~\ref{geom}. Numerical solution of this dispersion
equation shows that there exists a low frequency stop band for
all propagation directions (except for the particular case of
$q_z= k$ that will be analyzed latter). Let us simplify the
dispersion equation \r{disp} for the quasi-static case $a,b\ll
\pi/k$. Using the Taylor expansion of $\sin(x)$ and $\cos(x)$
functions for small arguments we obtain \r{qq} with
\e
k_0^2=\frac{2\pi/s^2}{\displaystyle\log\frac{s}{2\pi r_0}+F(r)},
\lab{k0}
\f
where $s=\sqrt{ab}$, $r=a/b$ and
\e
F(r)= -\frac{1}{2}\log r+ \sum\limits_{n=1}^{+\infty}
\left(\displaystyle \frac{\mbox{coth}(\pi n
r)-1}{n}\right)+\frac{\pi r}{6} \lab{Ffun}. \f Therefore, we have
shown the suitability of the suggested approach for the
description of the wire medium, with $k_0$ given by \r{k0}.
Parameter $k_0$ here plays the role of the wave number
corresponding to the plasma frequency. More exactly, it indicates
the upper edge of the low frequency stop band. Naturally, $k_0$ as
a function of two lattice periods $a$ and $b$ is a symmetric
function: $k_0(a,b)=k_0(b,a)$. It means that function $F(r)$ has
the following property: $F(1/r)=F(r)$. For the commonly used case
of the square grid ($a=b$), $F(1)=0.5275$. Expression \r{k0}
looks similar to the approximate expressions for the plasma
frequency developed earlier in \cite{Brown,Rotmanps,King,Pendryw},
but for thin wires it is more accurate and takes into account the geometry of the
lattice. Notice that the dispersion equation \r{qq} for
extraordinary waves is indifferent to the direction of the wires
axis: the wave vector components $q_x,q_y,q_z$ enter into this
equation completely symmetrically. It means that, within the low
frequency stop band, the extraordinary wave decays with the same
decay factor along all directions in space. The same can be said
for propagation in the first frequency pass band. This isotropy of
the dispersion equation is rather surprising since the medium is
strongly anisotropic. However, it can be shown from the very
fundamental facts summarized in \r{total} and \r{disuni} by
assuming, as usual, that \r{epl} is valid for extraordinary waves
propagating in the $x-y$ plane.

It is possible to transit \r{eps} from the spectral domain $(\vec
q,\omega)$ to the physical domain $(\vec r, t)$. The following
non--local material equation can be derived from \r{eps} using the
double Fourier transform:
\e
\vec D(x,y,z)=\varepsilon_0 \vec E(x,y,z)+ \f $$
\frac{\varepsilon_0k_0^2 c}{2} \_z_0 \int\limits_{-\infty}^t
\int\limits_{z-c(t-t')}^{z+c(t-t')} E_z(x,y,z',t') dz'dt', $$
where $c=1/\sqrt{\varepsilon_0\mu_0}$ is the speed of light in the
host matrix. Here, the area of integration in the $z-t$ plane is
the light cone $|z-z'|<c(t-t')$. In other words, the kernel in the
Fourier convolution is $u[c(t-t')-|z-z'|]$, where $u(x)$ is the
Heaviside step function. It means that the point $(x,y,z)$ inside
the wire medium (described as a dispersive continuum) at moment
$t$ is affected by the $z$-components of electric fields coming
from the domain $(x,y,z\pm c(t-t'))$ surrounding (along the wire
axis) this point during all the past time ($t'<t$). This result
illustrates the consistency of the reported model from the
relativistic standpoint.

In the following we will describe some relevant effects in the
analyzed parallel wire medium, associated with the spatial
dispersion. Refraction and reflection of plane waves at a plane
interface show strong differences between the local and non--local
models. Let us consider an interface between an isotropic
dielectric with the permittivity $\varepsilon_1$ and a uniaxial
dielectric with $\varepsilon$ described by \r{epl}. The interface
is in the $(y-z)$-plane and it is illuminated by a plane wave
coming from the isotropic dielectric. The wave vector and electric
field vector lie in $(x-z)$ plane ($q_y=0$, $E_y=0$). The
incidence angle of the plane wave is $\theta$. If
$\varepsilon_1>\varepsilon_0$, $\varepsilon<0$, and
$\sin^2(\theta)<\varepsilon_0/\varepsilon_1$ the wave will be
completely reflected, but for
$\sin^2(\theta)>\varepsilon_0/\varepsilon_1$ some part of the wave
will be transmitted through the interface. This transmitted wave
will be an extraordinary wave, as it follows from its polarization
state, and can be excited at any frequency. This amazing behavior
disappears when the non--local model summarized in \r{eps} is
used. Indeed, if the non--local axial permittivity \r{eps} is used
for the wire medium, we observe that no transmission inside the
wire medium is possible for $k<k_0$. At $k=k_0$ transmission is
possible only in the case of the normal incidence. Only if
$k>k_0$, a refracted wave appears.

\begin{figure}[h]
\centering \epsfig{file=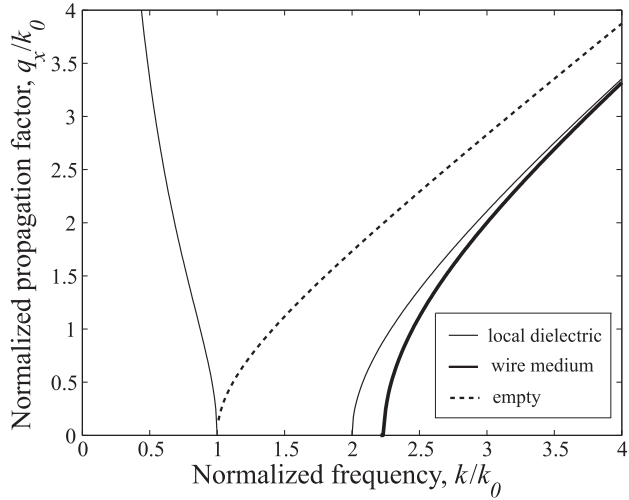, width=8.5cm} \caption{Normalized
propagation factors in a parallel--plate waveguide vs. normalized
frequency for different types of waveguide filling: 
thin solid lines -- uniaxial dielectric with a negative permittivity;
thick solid line -- wire medium; dashed line -- empty waveguide.
The wire medium and the uniaxial dielectric have the
same $k_0=\pi/(2d)$.} \label{prop}
\end{figure}

Let us next consider the guidance of electromagnetic waves in a
parallel-plate waveguide infinite in the $x$ and $y$ directions
and bounded by parallel perfectly conducting planes orthogonal to
the $z$-axis. Separation between the conducting walls is $d$. We
assume that this waveguide is filled with a wire medium with the
wires along the $z$-direction. We will consider eigenwave
propagation along the $x$ axis of the TM$_{01}$ mode
($H_y,E_x,E_z\neq 0$). For waveguides filled by a local uniaxial
dielectric with anisotropy axis along the $z$ direction we have
from \r{disuni}
\e
\varepsilon_0 q_x^2=\varepsilon (k^2-q_z^2),\quad
q_x=\sqrt{\frac{\varepsilon}{\varepsilon_0}
\left[k^2-\left(\frac{\pi}{d}\right)^2\right]}. \lab{propuniy} \f If
$\varepsilon>0$ Equation~\r{propuniy} gives a cut-off for
$k<\pi/d$ and propagation for $k>\pi/d$. In contrast, if
$\varepsilon<0$, propagation is allowed when $k<\pi/d$ (and
forbidden for $k>\pi/d$). Within this pass band a backward wave
($dq/d\omega<0$) propagates, as one can see in Fig.~\ref{prop}
(thin solid lines). 

This amazing behavior disappears if one fills
the waveguide with the analyzed non--local wire medium. Using
\r{qq}, we have in this case
\e
q_x^2+q_z^2=k^2-k_0^2,\quad
q_x=\sqrt{k^2-k_0^2-\left(\frac{\pi}{d}\right)^2}. \lab{propunix} \f
and we obtain the usual frequency behavior:  cut-off for
$k<\sqrt{(\pi/d)^2+k_0^2}$ and propagation for
$k>\sqrt{(\pi/d)^2+k_0^2}$. An increase of the cut-off frequency
is observed compared to the case when there is no filling medium,
as one can see in Fig.~\ref{prop} (thick solid line and dashed line).

Let us finally analyze the propagation of plane waves along the
wire medium for the particular case of $q_z=k$. In this
case, the non-local permittivity along the $z$-axis \r{eps}
becomes infinite. To avoid the singularity problem we use material equation of the form ${\bf E}=\=\varepsilon^{-1}{\bf D}$.
In this case the Maxwell equations have plane wave solutions for all frequencies.
For these waves the transverse wave-vector ${\bf q}_{\bot}= (q_x,q_y)^{\rm T}$ is arbitrary.
The waves are transverse with respect to the wire axis: $H_z=0$ and $E_z =0$.
The electric field is parallel to the transverse wave vector, ${\bf q}_{\bot}\times{\bf E}=0$.

Such waves can be interpreted as transmission-line
modes propagating along the parallel wires. In fact, a set
of $N$ infinite parallel wires can be viewed as a system of
coupled transmission lines. This system can support $N$ 
degenerate transmission line waves with $H_z=0$, $E_z=0$ and phase
constant $q_z=k$. The electric field of these waves can be
obtained from 
\e
{\bf E} = -({\bf u_x}\partial_x +{\bf
u_y}\partial_y)\phi(x,y)\exp{(-jkz)},
\f where $\phi(x,y)$ is a
quasi-electrostatic potential taking constant but arbitrary
values $\phi_n$ ($n=1,2,...,N$) at each wire. In fact, the plane wave
with transverse wave-number $\bf q_{\bot}$ and $q_z=k$ corresponds to the
transmission-line wave with $\phi_n=\phi_0\exp{(-j{\bf
q}_{\bot}\cdot{\bf r_n})}$, where ${\bf r_n}=(x_n,y_n)^{\rm T}$ is the
location of the $n$-th wire in the transverse $(x-y)$ plane. 

In summary, it has been shown that parallel wire media possess
very strong spatial dispersion effects at any frequencies,
including the very large wavelength limit. An analytical model for
the non--local permittivity dyadic of these media has been
presented and discussed. Inconsistency of the local model for
parallel wire media with non-vanishing wave-vector component along
the wires has been shown. Dramatic differences in the predicted
behavior of that media, arising from the use of the conventional local and/or the
non--local model for the permittivity are shown. Finally, the
proposed non-local model for the permittivity has been found to be
also suitable for the description of the transmission-line modes
of the structure. We feel that the reported results opens the
question of the role of spatial dispersion in the adequate
characterization of discrete metamaterials as effective media, at
least if arbitrary directions of propagation and/or polarization
of the electromagnetic field should be considered in the analysis.
In addition, an example has been presented of an effective
medium in which spatial dispersion is important at any frequency,
in contrast with some commonly assumed ideas about the physical
relevance of this effect.

\bibliography{spdisp}

\end{document}